\documentclass[amstex,epsf,amssymb,usenatbib]{mn2e}
\usepackage{epsf}
\usepackage{amsmath,amssymb}
\title[Simulations of outflows]
{Simulations of Evolving or Outbursting Molecular Protostellar Jets}

\author[A. Rosen \& M.D. Smith]
       {Alexander Rosen\thanks{E-mail: rar@star.arm.ac.uk}
\& Michael D. Smith\thanks{E-mail: mds@star.arm.ac.uk} \\
Armagh Observatory, College Hill, Armagh BT61 9DG, Northern Ireland \\
}

\date{Accepted .....
      Received ..... ;
      in original form .....}

\pagerange{\pageref{firstpage}--\pageref{lastpage}}
\pubyear{2003}

\begin{document}

\maketitle

\label{firstpage}

\begin{abstract}
The kinematic and radiative power of molecular jets is expected to change as a protostar
undergoes permanent or episodal changes in the rate at which it accretes.
We study here the consequences of evolving jet power on the spatial and
velocity structure, as well as the fluxes, of molecular emission from
the bipolar outflow. We consider a jet of rapidly increasing density and
a jet in which the mass input is abruptly cut off. We perform three
dimensional hydrodynamic simulations with atomic and molecular cooling and
chemistry. In this work, highly collimated and sheared jets are assumed. We find that
position-velocity diagrams, velocity-channel maps and the
relative H$_2$ and CO fluxes are potentially the best indicators of
the evolutionary stage. In particular, the velocity width of the CO
lines may prove most reliable although the often-quoted mass-velocity 
power-law index is probably not. We demonstrate how the relative 
H$_2$~1--0~S(1) and CO~J=1--0 fluxes evolve and apply this to
interpret the phase of several outflows. 
 
 \end{abstract}


\section{Introduction}              

Jets are being increasingly associated with protostars and with bipolar 
molecular outflows.  Prominent examples of jet-driven collimated outflows include 
HH\,111 \citep{1997ApJ...482L.195N,2001ApJ...559L.157H},
HH\,211 \citep{1999A&A...343..571G, 2001ApJ...555..139C},
HH\,212 \citep{2000ApJ...542..925L} and
HH\,288 \citep{2001A&A...375.1018G}.
For many other outflows, there is  evidence that their momentum is also derived 
from the thrust of jets. For example, transverse shock structures behind bow 
shocks are interpreted as the sites of jet impact \citep{skd03}. 
Nevertheless, estimated jet thrusts appear insufficient to
drive some outflows \citep{1998ApJ...499L..75F} and
ballistic bullets and wide winds have been invoked. 
Alternatively, as investigated here, the presently observed jet may not reflect 
the ancestor jet that has previously driven the outflow.\\

Our goal is to determine the observable characteristics of outflows
driven by molecular jets with evolving mass flow rates.
Jet evolution was considered as an interpretation of the systematic
decrease in flow velocity with distance in the atomic HH\,34 outflow
\citep{2000A&A...354..667C}. The
deduced requirements appeared unlikely and an alternative 
precessing-fragmenting scenario was favoured. A systematically increasing
jet speed was simulated by \citet{2002MNRAS.335..817L}. Their idea was to explain how molecules
can be accelerated to high speed without destruction. Here, we choose to evolve the
mass flow rate rather than the speed since it is clear that the mass
flow rate must change by many orders of magnitude as a protostar evolves.
The taxonomy of protostars is defined by the categories Class 0 through 
Class 3, which have been linked to an evolutionary sequence 
\citep{1987IAUS..115....1L, 1993ApJ...406..122A}.  One scenario connects this
sequence to simultaneous evolutions in both the accretion rate and the outflow 
rate \citep{2000IrAJ...27...25S}. Furthermore, certain protostars undergo 
outbursts \citep{1996ApJ...466..844N}, the consequences of which may  also show 
up later in the outflow properties \citep{2001ApJ...551L.171A}.

We investigate a limited set of three-dimensional
simulations that correspond to significant and permanent changes to the input mass flux.  
This complements studies of outflow development generated by non-evolving 
heavy molecular jets \citep{1997A&A...318..595S,1998MNRAS.299..825D,1999A&A...343..953V}
as well as a range of jet conditions \citep{rs03}.
We first consider a rapid decrease in the mass flux, which we
suspected would produce a bullet-like outflow.
A decrease in mass flux of some variety must accompany the
transitional stages of a protostar into a star.
In addition, we model a jet that increases density after some initial
propagation, and this model correlates with the opposite end of a protostar's evolution.
During the initial collapse of the core into a protostellar
disk, a very young jet must ignite and, therefore,
be characterised by an increase in mass flux.

\begin{figure*}
  \begin{center}
  \epsfxsize=15.0cm
    \epsfbox[10 10 440 260]{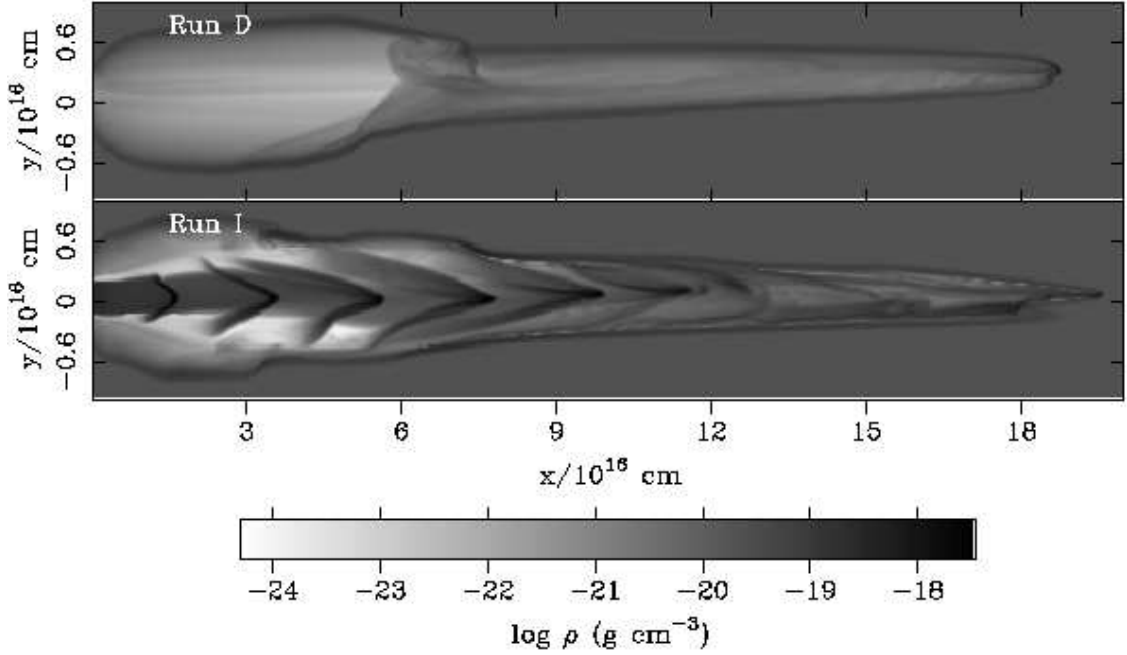}
\caption[]
{Midplane cross sections of density.  These are cross sections for z/R$_{j}$ = 0.0 of
Run D at t = 1300 yr (top) and Run I at t = 1400 yr (bottom). Each is scaled to the
same maximum and minimum and the scale, with darker shading indicating denser regions,
is displayed below the cross sections. The complete computational domain in the two
axes is displayed. In both panels, the vertical axis is the $y$-axis.}
\label{bothmid}
  \end{center}
\end{figure*}

Certain features of the molecular line emission from outflows have been
attributed to an evolving jet mass outflow rate. For example, it is
believed that the energy radiated from shock waves
provides a measure of the present driving power while the mechanical
luminosity provides a record of the driving history. In this way,
the ratio of  emissions from vibrationally excited molecular hydrogen
and rotationally excited carbon monoxide could determine the
evolutionary phase of the protostellar exhausts 
\citep{1998MNRAS.299..825D,2000AJ....120.1974Y}.
We can determine here the validity of this statement, although these simulations are
of a particular type and scale and so only begin to illustrate the
full range of possibilities. For example, we take here a hydrodynamic
highly-collimated jet and a uniform ambient medium that are both 
initially fully molecular. In the next section, we discuss more fully the
characteristics of the jet simulations.

\section{ZEUS-3D with Molecular Cooling and Chemistry}

\subsection{The code}

We have modified the ZEUS-3D code, which usually updates variables with an explicit method, 
to include a semi-implicit method that calculates the molecular and atomic hydrogen fractions,
according to the prescription of \cite{1997A&A...318..595S}. We have also added a limited 
equilibrium C and O chemistry to calculate the CO, OH and H$_2$O abundances \citep{rs03}.
Equilibrium CO and H$_2$O is a reasonable estimate at high density, consistent with the 
low numerical resolution of the cooling layers behind shock waves \citep{rs03}. 

The details of the many components of the cooling function and chemistry network 
are discussed in the appendices of \citet{sr03}.  As an overview, we include cooling
through rotational and vibrational transitions of  H$_2$, CO, and H$_2$O, H$_2$ dissociative 
cooling and reformation heating, gas-grain cooling/heating, and a time-independent atomic 
cooling function that includes non-equilibrium effects. We include an 
additional 10\% by number of helium atoms, so the mean particle mass is 2.32 $\times$
10$^{-24}$ g. The dust temperature is fixed at 20\,K.

The ZEUS code suffers from numerical errors. Recently, \citet{2002ApJ...577L.123F}
has explored the appearance of rarefaction shocks which, as the name suggests,
are particularly resistant and their influence cannot be removed by increasing the 
resolution. They may occur in locations where the local Courant number is near 0.5
and where the velocity is negative. Inaccurate fluid patterns could result for
adiabatic or MHD flows but isothermal flows are not significantly in error. The
present jet simulations are thus immune to these errors, with strong forward motion
and strong cooling, especially in the zones with relatively large Courant number (i.e. approaching
0.5). In any case, isothermal flows are an approximation in which regions undergoing rarefaction are
being artificially heated (to counter expansion cooling). Cooling also generates
structure through thermal instabilities on many scales which are not adequately 
featured in the numerical approximation. Nevertheless, we persist with the ZEUS code
because of its high versatility, allowing the introduction of new physics, the
discovery and removal of specific errors over many years, and its speed of
execution.   

\subsection{The fixed conditions}

Two models for the time evolution of the jet are studied. We designate 
these as Simulations D and I, with a decreasing and increasing jet density,
respectively.  

In both 3D simulations, the computational grid spans 2~$\times$~10$^{17}$~cm 
along the jet axis ($x$) and 2~$\times$~10$^{16}$~cm in both transverse 
dimensions, with 1000 $\times$ 100 $\times$ 100 uniform zones in each direction.  
We initialise the jet with a nozzle radius, R$_j$, of 1.7 $\times$ 10$^{15}$ cm 
or 7.5 zones.  The initial internal Mach 
number of the jets is 140 with a temperature for the fully molecular jets of 100 K; 
this combination gives a speed of 100 km s$^{-1}$. The ambient medium is
molecular with a hydrogen nucleus density of 10$^4$~cm$^{-3}$.

In addition, both jets are pulsed with an amplitude of $\pm$ 30\% and a period 
of 60 yr, with a $\sin (\omega t)$ dependence. A small precession
with a 1$^{\circ}$ half-precession angle and a period of 50 years
is included. A radial shear that reduces the speed to 70\% of the core velocity at 
the jet edge is also included. The sinusoidal form of the pulsation and the 
parabolic form of the shear follow Eqns. 1 and 2 of \citet{1999A&A...343..953V}.

\subsection{The evolving parameters}

Two models for the time evolution of the jet are studied. We designate 
these as Simulations D and I, with a decreasing and increasing jet density,
respectively. 

In Simulation D, we decrease the jet density abruptly,
setting it to zero after a short period of time.  We resorted to this dramatic
variation of the flow because earlier attempts with more moderate decreases
failed to show a significant deceleration.  The end of the run may thus correspond
to a highly evolved jet-driven outflow.
Specifically, we allow the ``jet" to propagate
for a mere 20 years and then sever its ties to the life-giving momentum source
at the inlet.  
The jet is initialised with a hydrogenic nucleon density
of 10$^5$~cm$^{-3}$ ($\log n_j$ = 5).  In addition we have simulated other
aborted jets, with flows terminated after 20 years but the flow was not 
initially precessed,
and also a case where the flow was terminated after only 5 years
and also not initially precessed.

\begin{figure*}
  \begin{center}
  \epsfxsize=18.0cm
    \epsfbox[10 10 490 230]{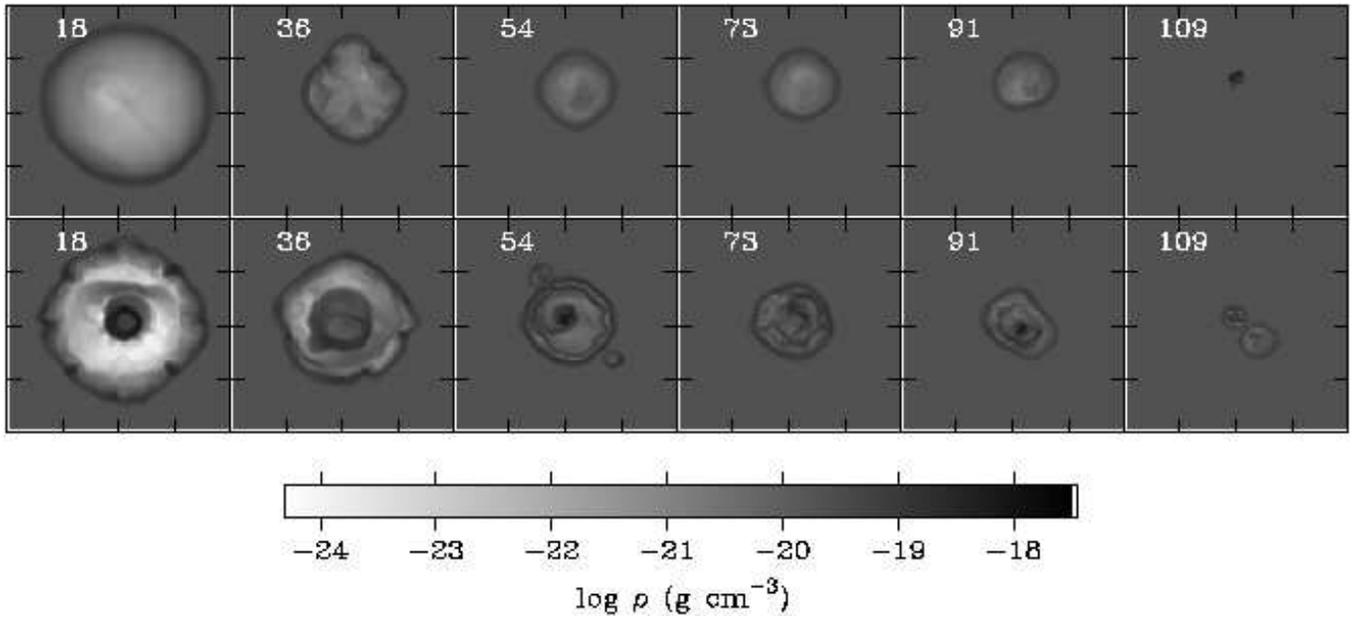}
\caption[]
{ Axial cross sections of density from Run D (upper row) and from Run I (lower row).
Each row is from the time displayed in the corresponding panel in Figure
\ref{bothmid}.  Each cross section is scaled to the same maximum and minimum and the scale,
with darker shading to indicate denser regions, is displayed below the cross sections.
We display the complete computational domain in the two axes shown.  The distance between
adjacent tick marks is 5 $\times$ 10$^{15}$ cm. The axial position in R$_{j}$ is given in
the upper left of each panel. The vertical axis is the $y$-axis, and the
horizontal axis is the $z$-axis, and so the view is from the jet inlet boundary looking down
the jet axis (i.e., toward the +$x$-direction).  }
\label{bothax}
  \end{center}
\end{figure*}

In Simulation I, we wish to show the consequences of increasing the mass density 
in a light jet flow that had already propagated some distance.  
In this case, the density evolution follows a less abrupt change, determined by 
the following:
\begin{equation}
\rho_j(t) = \rho_0 + \rho_1 \tanh \Big(\frac{t - t_1}{t_0}\Big)
\end{equation}
where we set $\rho_1$ = 1.1484 $\times$ 10$^{-19}$ g cm$^{-3}$, $\rho_0$ = 1.1716
$\times$ 10$^{-19}$ g cm$^{-3}$, $t_1$ = 800 years, and $t_0$ = 50 years.  This results
in the logarithm of the jet hydrogenic number density rising from $\log n_j$ = 3 to 
$\log n_j$ = 5 over a 200 year period centered at a program time 800 years after the 
simulation has started.  Given the axial grid length of 2 $\times$ 10$^{17}$ cm, 
this leads to the bow shock from the initially light jet
being roughly half way across the grid when the mass flux is increased.

\section{Internal Structure}

Slices of jet density down the midplane are presented in Figure \ref{bothmid}.  Both
simulations feature a refocused leading region and wider trailing ``shoulders".
We remark that the shoulder regions in both simulations shown in Fig.~\ref{bothmid}
have progressed to only 7~$\times$~10$^{16}$~cm, a small fraction of the jet length.
In our study of non-evolving jets \citep{rs03}, we were able to
attribute the refocusing to the combination of both shear and strong molecular cooling,
since simulations with atomic jets or molecular jets with no shear do not refocus. 
The ratio of the distance of the shoulders from the nozzle relative to that of the 
entire jet was found to be proportional to the jet-to-ambient  density ratio.
Hence, without a continuous strong power source in Simulation D, the bloated cocoon is stunted.

Furthermore, the average speed of the leading bow shock, which almost traverses the grid in
1300--1400 years, is $\sim$ 45~km~s$^{-1}$.
This is similar to a ``light" jet simulation (jet-to-ambient density ratio of 0.1 in
\citealt{rs03}).  This average speed is still, however,
nearly 90\% faster than expected from ram pressure arguments due to the concentration of the jet
thrust caused by the refocusing.

\begin{figure}
  \begin{center}
  \epsfxsize=8.0cm
 \epsfbox[55 360 280 530]{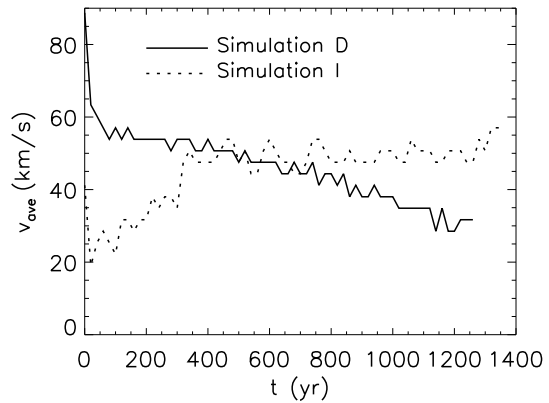}
\caption[]
{ Time-averaged advance speed of the bow shock.  The advance speed here is determined
from the leading edge of the bow shock in integrations of H$_2$ emission, similar to Figures \ref{int1}
and \ref{int2}. }
\label{avel}
  \end{center}
\end{figure}

From a plot of a time-averaged advance speed of the bow shock (Figure \ref{avel}), we see that
the advance speed of the bow shock is not constant throughout either simulation.
In Run D, the advance speed of the jet falls below 60 km~s$^{-1}$ within the first 100 years,
and follows a linear decrease in velocity with time until t $\sim$ 800 years, where
it continues a linear decrease but at a faster rate.  The change of the deceleration rate is
related to a broadening of the tip of the bow shock, which is at the leading edge of a
region of refocused and accelerated flow after $t \sim$ 200 years.  At the end of
Simulation D the bow shock is progressing at a speed of $\sim$ 30 km s$^{-1}$.  This is an
example of the ability of all of our aborted molecular jet simulations to maintain its 
momentum for the length of the computational grid (0.06 pc).  Even in the case of the jet 
aborted after 5 years, we still found a speed of $\sim$ 20--25 km s$^{-1}$ after the 2000 year
interval in which it crossed the grid.

In contrast, the progress of the bow shock in Simulation I, as measured from the molecular 
emission maps presented below, increases early in the simulation.  The speed initially drops 
to 20 km~s$^{-1}$, but after only 50 years begins to rise.  There is a significant increase 
at $t \sim$ 300 years, after which the speed
of advance of the bow shock is $\sim$ 50 km~s$^{-1}$.  The increase does not
coincide with the onset of the higher jet density but, instead, coincides with the initial
refocusing of the bow shock at its leading edge.  Thus, the
stripping of the low-speed jet sheath is an
interesting means by which an acceleration in the proper motions of jet knots
may result. Even after the jet density has increased, the average advance speed for the 
bow shock remains nearly constant until the simulation is terminated. 

\begin{figure*}
  \begin{center}
  \epsfxsize=17.5cm
 \epsfbox[10 10 490 230]{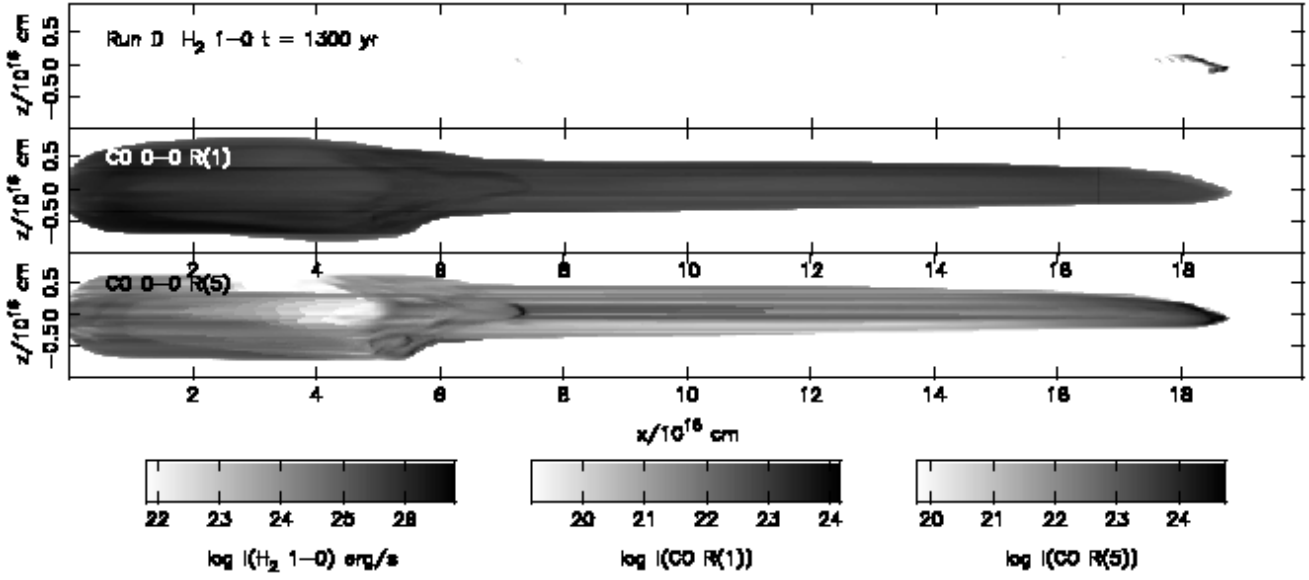}
\caption[]
{ Appearance of Simulation D in three molecular emission lines.
The luminosities from each zone are placed in bins in the viewing window that are
the same size as the 3D zones used in the simulations (i.e., 2 $\times$ 10$^{14}$ cm).
The axis being integrated is the $z$-axis, so the vertical axis in all the panels is the $y$-axis.}
\label{int1}
  \end{center}
\end{figure*}

\begin{figure*}
  \begin{center}
 \epsfxsize=17.5cm
    \epsfbox[10 10 490 270]{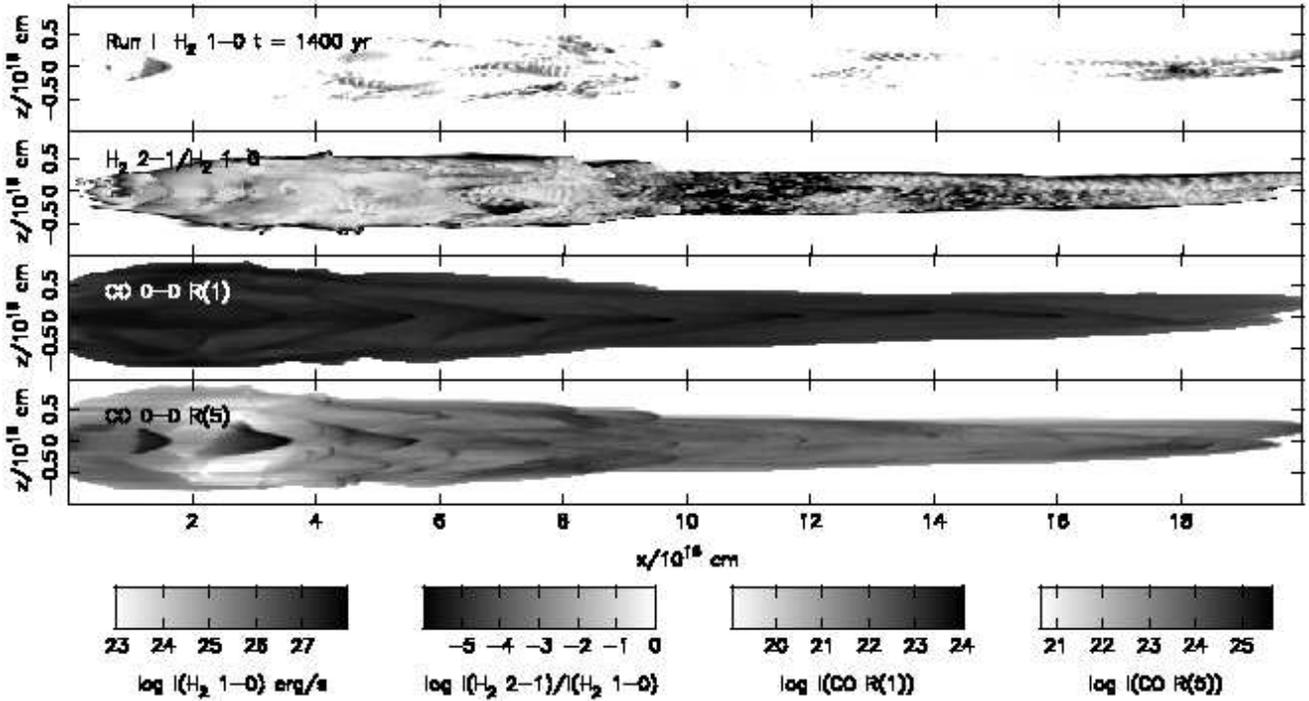}
\caption[]
{ Appearance of Simulation I in four molecular emission lines.  See caption
in Fig.~\ref{int1}.  Since this simulation generated sufficient amounts of warm (T $\sim$ 10,000 K) gas, we show
the map of the H$_2$ 2--1 S(1) emission in the second panel.}
\label{int2}
  \end{center}
\end{figure*}

The dense material in Simulation D is confined to the bow shock and swept-up shell
while dense material is also associated with the internal shocks of the pulsed jet in Simulation I.
The cocoon surrounding the jet behind the shoulder region has the lowest density, with the 
still-connected jet in Simulation I having a lower minimum density by a 
couple of orders of magnitude than the aborted jet in Run D.  These overdense and underdense 
regions within the jet are also apparent in the axial cross sections of density in 
Figure \ref{bothax}.  With the ability of the jet in Run I
to generate a very low density cocoon, the relatively small overdensity (still an order of 
magnitude) in the bow shock is difficult to see. 
Note twin concentric shell structures are formed in the cross section of Run I,
visible in  Fig.~\ref{bothax} at distances R$_j$ = 54--91, where the 
powerful jet transmits invigorated shocks through  the cocoon. 

\section{Simulated Molecular Line Emission}

\subsection{The Spatial Distributions}

Images of line emission from molecules are shown for Simulation D in Fig.~\ref{int1} and for 
Simulation I in Fig.~\ref{int2}.  Integration is across the $y$-direction
transverse to the jet axis, and the simulation time corresponds to the previous figures.
The H$_2$ line emission is based on a non-LTE approximation to the vibrational populations 
\citep{1979ApJS...41..555H}, and the CO emission is calculated from the formulae of 
\cite{1982ApJ...259..647M}. In addition,  we set a lower velocity limit of 1 km~s$^{-1}$ 
on the integration of CO emission lines in order to subtract out the undisturbed cloud.

In Figure \ref{int1}, we show the H$_2$ 1--0 S(1) line, the CO R(1) and R(5) lines
for the CO ground vibrational state ($\upsilon$ = 0).  In the H$_2$ line, only the leading edge of the
bow shock remains visible and, since Simulation D is  relatively devoid 
of warm (T $>$ 5000 K) gas, the image in the H$_2$ 2--1 S(1) line is not shown.
We display instead the CO R(5) line, which should highlight gas at intermediate temperatures
(T $\sim$ 100 K) between the H$_2$ line and the CO R(1) line (which emphasizes the coldest,
T $\sim$ 5 K gas).  The CO R(5) line image not only has an intermediate appearance 
but also shows some filamentary structure, which is brightest
at the leading edge of both the bow shock and the shoulders. This relatively
warm gas is generated by the accelerating refocused region.  In both of the simulations
presented here, the moving CO R(1) shows the outline of the bow shock.

\begin{figure*}
  \begin{center}
  \epsfxsize=15.0cm
 \epsfbox[50 360 525 650]{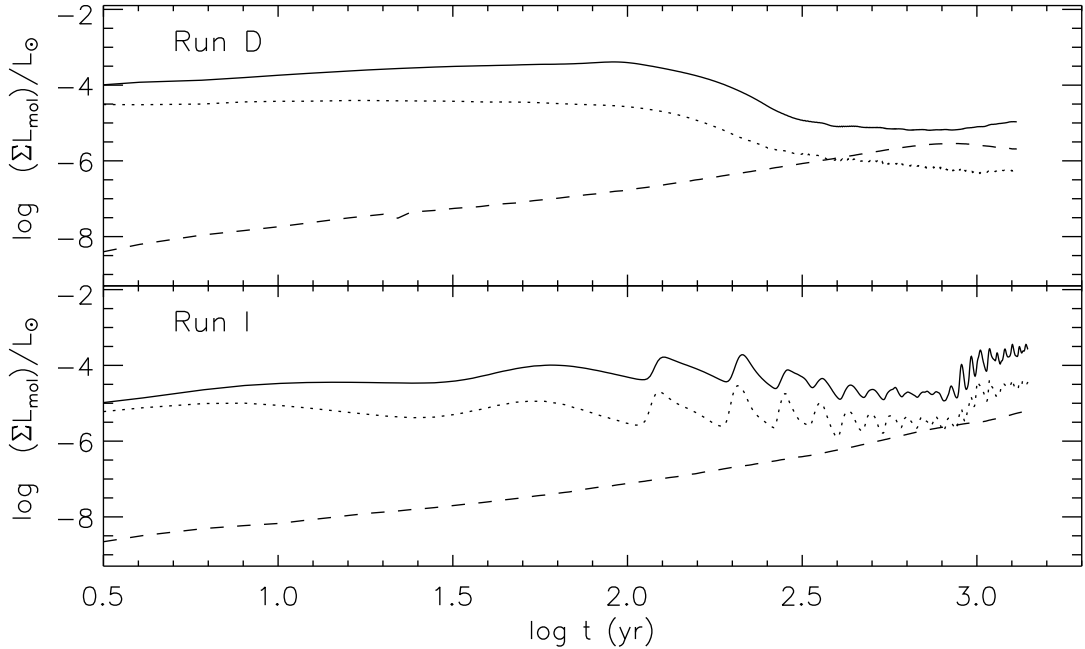}
\caption[]
{ Integrated molecular emission in three lines for each simulation. The H$_2$ 1--0
emission is the solid line, the H$_2$ 2--1 is the dotted line, and the CO emission
is the dashed line. }
\label{allrad}
  \end{center}
\end{figure*}

Simulation I is sufficiently energetic to generate copious amounts of 10,000 K gas.
Hence, in addition to the lines shown in Fig.~\ref{int1}, we also show the relative
luminosity of the H$_2$ 2--1 line to the 1--0 line in
Fig.~\ref{int2}.  The emission in the H$_2$ 1--0 image is sparse
and filamentary, showing several `minibows' especially at the edges of the
shoulder region near $x =$ 9 $\times$ 10$^{16}$ cm.  Only a single pulse near 
the inlet is seen in the H$_2$ 1--0 map. Earlier pulses, farther down
the jet, are easier to see in the ratio of the H$_2$ lines, and are noticeable in 
the CO R(1) line, but very prominent in the map for the CO R(5) line. While the 
CO R(1) emission map does show a wider shoulder region well behind the leading 
edge of the bow, this region is much less pronounced than the appearance  in the
midplane density slice.

\subsection{Totals}  

The evolution of the totals of the H$_2$ emission lines and the CO R(1) line are shown
in Fig.~\ref{allrad}.  In Simulation D, a dramatic reduction in the H$_2$ totals begins after
$t$ = 100 yr. A flattening of the CO luminosity, which monotonically increases for non-evolving
molecular jets, also occurs but not until $t$ = 700 yr. Thus, a significant change in the 
ratio of the H$_2$/CO total emission occurs shortly after the flow has been turned off. 
It is only when the H$_2$ 1--0/CO R(1) ratio approaches unity that the total
emission from each emission line tends to flatten out. Before this occurs the ratio is a 
good age indicator for the aborted jet flow.

The 80 year delay between the jet demise and the H$_2$ weakening is clearly the time to 
elapse before the leading bow stops being driven by the jet impact. The 700 year delay 
between the termination of the jet flow and the flattening of CO R(1) emission
indicates that new ambient material continues to be snowploughed, as the momentum of a 
small fraction of high-speed gas is redistributed into a large fraction of low-speed gas. 
This is a true `bullet' or ballistic phase. To test this, the deceleration of the bow 
shock in Run D as shown in Fig.~\ref{avel} can be compared to the analytic motion of a
bullet of speed $v_b$, fixed mass M, cross-sectional area A, and initial density $\rho_{bo}$, 
moving into a medium of density  $\rho_a$, with a drag coefficient D$_c$. The equation 
of motion is 
\begin{equation}
   \frac{d\,v_b}{d\,t} = - \frac{v_b^2}{L}
\end{equation}
where $L =  M/({\rm D}_c\rho_aA)$. Here L defines an effective column length of ambient 
gas required to slow down the bullet. First, we take ram pressure confinement of a bullet 
which maintains its shape as the pressure decreases. This yields
\begin{equation}
   L = L_o \left[\frac{\rho_{bo}}{\rho_{a}}\right]^{2/3}\left[\frac{v_b}{c_s}\right]^{4/3},
\end{equation}
where $c_s$ is the bullet sound speed and $L_o$ is the initial value of L. Substitution 
and integration gives
\begin{equation}
   t = 3\frac{L_o}{v_{bo}}\left[1 - \left(\frac{v_b}{v_{bo}}\right)^{1/3}\right]
\end{equation}
and
\begin{equation}
   x = \frac{3}{4}\,L_o \left[1 - \left(\frac{v_b}{v_{bo}}\right)^{4/3}\right].
\end{equation}
Alternatively, if the bullet is either disk shaped or does not expand at all, L is a 
constant and we find 
\begin{equation}
   \frac{v_b}{v_{bo}} = \frac{1}{1 +  t\,(v_{bo}/L)}.
\end {equation}
For the particular conditions of Run D, $L_o = (6 \times 10^{16}/D_c)$~cm.
To decelerate from 60~km~s$^{-1}$ to  30~km~s$^{-1}$ in 1200~yr, then
requires D$_c$ = 0.17--0.28, depending on the bullet model utilised.
This drag coefficient is then consistent with the aerodynamic shape of the leading bow
and thus shows that a large mass can be mildly disturbed by a high-speed jet.

The duration of this phase is 35 times longer than the jet-driving phase, suggesting
an effective jet column density which exceeds the deflected ambient column density by 35.
We can roughly attribute this to the high jet density (10) and the refocusing (3.5).

It should be remarked that decelerating bullets have been investigated in various
astrophysical contexts including Herbig Haro objects \citep{1979ApJ...228..197N},
radio galaxies \citep{1980A&A....81..282S} and planetary nebula
\citep{1996A&A...307..225P}. Depending on how the bullet forms, 
stability is maintained through inertial confinement for a sound-crossing time 
or a shock-wave crossing time. Therefore, the high Mach number of the bullet in Run D 
permits a bullet to travel over 10$^{17}$~cm without great expansion.  

In Run I, periodic brightening in H$_2$ luminosity occurs as the pulses in
the flow pass through the shoulder region and energise the bow shock envelope. 
The increase in jet density that becomes
significant for $t >$ 750 yr shows up as an increase in the H$_2$ output very
soon thereafter ($t$ = 800 yr).  No accompanying increase in the CO emission is
noticeable. There is an unrelated early break in the rate of increase in
the CO R(1) total luminosity, at t = 100 yr.  The slightly increased rate is 
related to the shape of the envelope, which shows the effects of refocusing near
the start of the simulation.

The final integrated fluxes of the two simulations have contrasting behaviours. In Run D,
the absolute luminosity in the 1--0~S(1) line of H$_2$ is continued at the low level
of L(1--0~S(1)) $\sim$~10$^{-5}$~L$_\odot$ despite the lack of a jet. This compares to the
value of $\sim~10^{-3}$~L$_\odot$ for the equivalent simulation in which the jet
power is maintained \citep{rs03}.  In Run I, L(1--0~S(1)) reaches 3~$\sim~10^{-4}$~L$_\odot$
and is set to increase. It is clear that a complex relationship exists between the immediate
jet power and L(1--0~S(1)). Whereas we found this ratio to lie in the range 80--600 for
non-evolving jets \citep{rs03}, we here find values between zero (Run D, after 20 yr) and
1.2 $\times$ 10$^5$ (Run I, 800 yr).

The predicted values of L(1--0~S(1)) are comparable to those derived in the unbiased
statistical study of \citep{2002A&A...392..239S}, where
the majority of outflows possess L(1--0~S(1)) in the range 10$^{-4}$--10$^{-3}$~L$_\odot$.
Nevertheless, the absolute fluxes are quite low  in comparison to a number of 
observations of well-known outflows. This is partly due to the compactness of the 
simulated outflows, limited by our need to follow molecule dissociation. We can argue 
that scaling up the spatial dimensions by a factor of 10 would scale the luminosities 
by a factor of 100. It remains to be shown, however, that the flow patterns would 
be maintained.

\section{Velocity Distribution of Mass and Molecular Line Emission}  

\begin{table*}
     \caption[]{Mass Spectra Power-Law Dependences}
     \label{gammas}
     \begin{tabular}{lccclllcclll}
          \hline
          \noalign{\smallskip}
 {\rm Sim.} &  t(yr) & type  & $\theta$ & range of log v & $\gamma$ & N$^{\rm a}$ & type & $\theta$ & range
of log v & $\gamma$ &  N$^{\rm a}$\\ \noalign{\smallskip}
\hline
\noalign{\smallskip}
D & 1300 & mass & 15 & 0.5--1.0 & 1.19 & 7 & CO & 15 & 0.2--0.6/0.7--0.9 & 1.27/1.92 & 2/3 \\
            &         &         & 30 & 0.5--1.0 & 1.17 & 7 &      & 30 & 0.5--1.0/0.9--1.1 & 1.17/2.74 & 7/5 \\
            &         &         & 45 & 0.5--1.0 & 1.19 & 7 &      & 45 & 0.5--1.0/1.1--1.2 & 1.22/3.58 & 7/3 \\
            &         &         & 60 & 0.5--1.0 & 1.10 & 7 &      & 60 & 0.5--1.0/1.1--1.3 & 1.16/3.07 & 7/7 \\
            &         &         & 90 & 0.5--1.0 & 1.13 & 7 &      & 90 & 0.5--1.0/1.2--1.3 & 1.21/3.18 & 7/7 \\
             \noalign{\smallskip}
I        & 1300 & mass & 15 & 0.5--1.0 & 1.05 & 7 & CO & 15 & 0.5--1.0 & 1.20 & 7 \\
            &         &         & 30 & 0.5--1.0 & 1.01 & 7 &       & 30 & 0.5--1.0 & 1.22 & 7 \\
            &         &         & 45 & 0.5--1.0 & 0.99 & 7 &       & 45 & 0.5--1.0 & 1.22 & 7 \\
           &         &         & 60 & 0.5--1.0 & 1.01 & 7 &       & 60 & 0.5--1.0 & 1.21 & 7 \\
           &         &         & 90 & 0.5--1.0 & 1.03 & 7 &       & 90 & 0.5--1.0 & 1.21 & 7 \\
\noalign{\smallskip}
         \hline
    \end{tabular}
\begin{list}{}{}
\item[$^{\rm a}$] N is the number of points in the velocity distribution, which was computed in 1 km s$^{-1}$ bins,
used to estimate the slope.
\end{list}
\end{table*}

\subsection{Mass ``Spectra"}   

CO velocity data for observed outflows are now invariably analysed by
plotting  the differential mass in velocity bins, with a few simple
assumptions made to compute mass from CO intensity.  One may classify how this mass
decreases with velocity by fitting a power law, with the slope defined by
$dm/dv \propto v^{-\gamma}$.  Two surveys containing many sources
(\citealt{2000AJ....120.1974Y} and \citealt{2001A&A...378..495R}) conclude that  a
broken power law is often necessary, with a shallow slope of 1.0--2.0
at low velocities and a steeper slope, in some cases $\gamma$ is close to 10, at
larger velocities.  The velocity where the break occurs is usually
$\sim$ 10 km s$^{-1}$.  One interpretation of this break speed is that it is the projected
component of the critical shock speed, $v_d\cos \theta$ with $v_d$ = 23 km s$^{-2}$
(e.g., \citealt{1994MNRAS.266..238S}), above which all molecules are assumed to dissociate
within the shock.  

\begin{figure*}
  \begin{center}
  \epsfxsize=15.0cm
 \epsfbox[10 10 500 350]{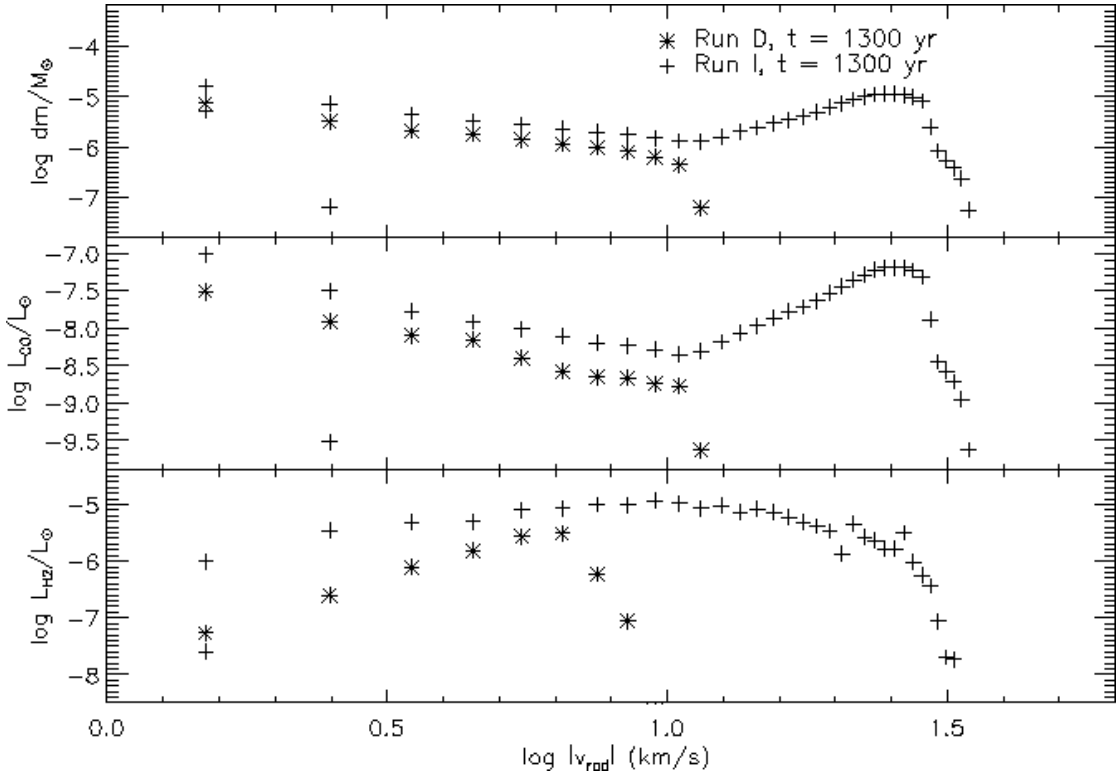}
\caption[]
{ Distribution of velocities into bins of mass and two molecular line luminosities for a viewing
angle of 15$^\circ$.  Displayed are the distribution of mass
(top panel), CO luminosity (middle panel) and H$_2$ 1--0 emission
(lower panel). Each velocity bin is 1 km~s$^{-1}$ wide.
The designation for the data presented in each panel is shown in the top panel.
Each run may be represented twice within each panel, with data from both positive and
negative radial velocities included.  Naturally, the smaller ranged data is for the
positive radial velocities (which could contribute to a  ``red" lobe, while the other 
is from a ``blue" lobe). }
\label{gamma}
  \end{center}
\end{figure*}

We have performed a similar analysis for the velocity distributions of mass,
of CO R(1) luminosity and of H$_2$ 1--0 luminosity for each of the aborted jet simulations,
including Simulation D, and for Simulation I.  In the case of Simulation D, the analysis was performed
at a program time of $t$ = 1300 yr, corresponding to other Simulation D illustrations
presented here.  In the increased-density jet of Simulation I, the tip of the jet has just 
barely reached the outer $x$-boundary at $t$ = 1400 yr. We, therefore, chose
to inspect data at $t$ = 1300 yr. We list the results for a sample of viewing angles,
defined as the angle of the jet axis out of the
plane of the sky and toward the observer, in Table \ref{gammas}.  We display the
data for one viewing angle (15$^{\circ}$) in Figure \ref{gamma}.

Having only found low values of $\gamma$ in previous simulations involving molecular
jets (although high values are possible for an atomic jet),
we have chosen to investigate whether a reduced mass flux
would give larger values for $\gamma$.
In Simulation D and the other aborted jet simulations, however, we still find shallow dependences
of the mass distribution on velocity i.e. small values for $\gamma$.  The values for
the aborted jet simulations are in general larger than their counterparts in
the increasing jet-density simulation, with the values for both sets usually nearer
1.0 than 2.0.   However, large slopes are computed for high velocities in the only cases
where a break could be seen, which are the CO-derived values for the aborted jet simulation.
These results are not significantly different than those found for non-evolving jets
\citep{rs03}.

\begin{figure*}
  \begin{center}
  \epsfxsize=15.0cm
 \epsfbox[10 10 490 240]{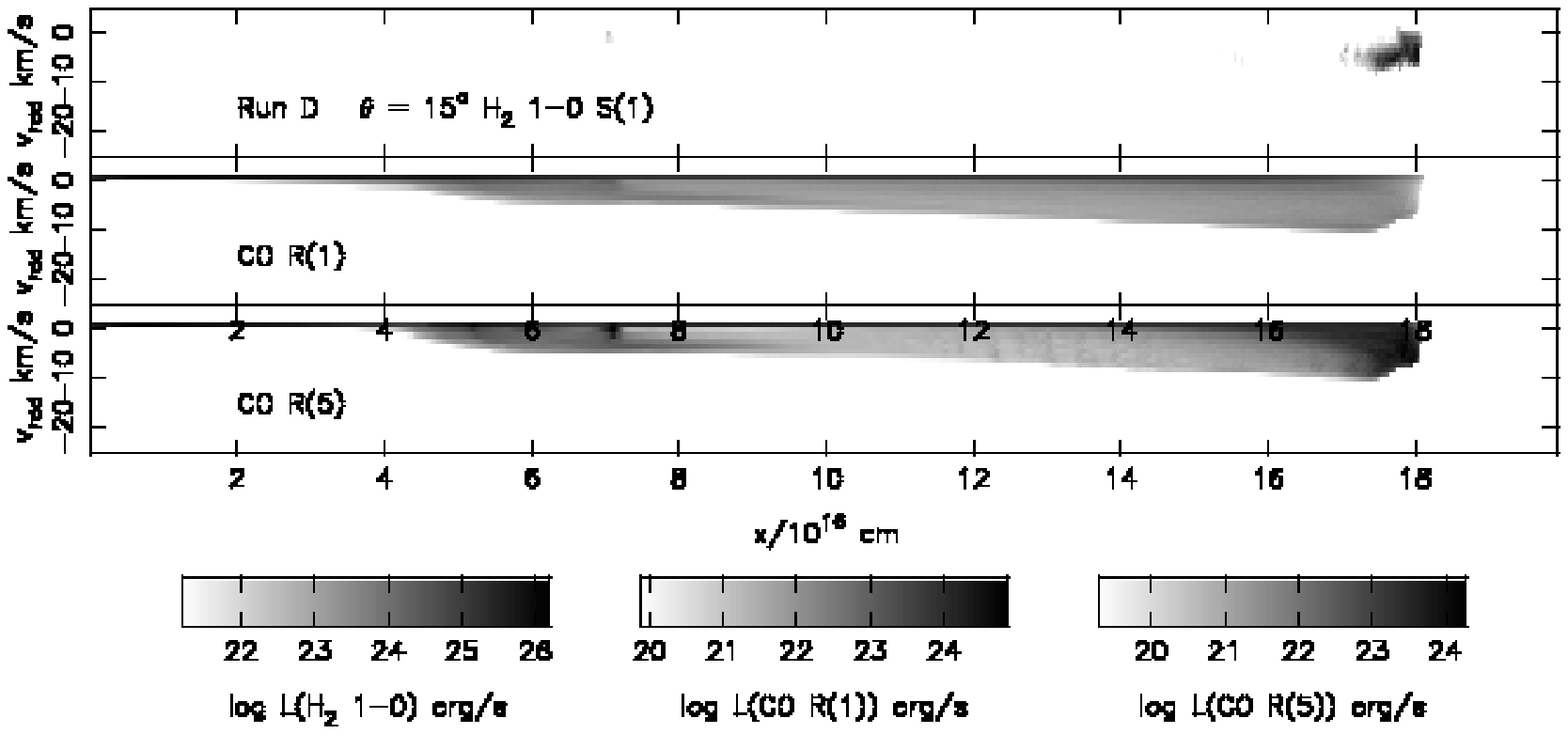}
\caption[]
{ Position-velocity map for Run D in three molecular emission lines.  The jet axis is 15$^{\circ}$
out of the plane of the sky toward the observer.  Each bin spans 2 $\times$ 10$^{14}$ cm in $x'$, the jet
axis projected on the plane of the sky, and 1 km~s$^{-1}$ in radial velocity. }
\label{posvel1}
  \end{center}
\end{figure*}

\begin{figure*}
  \begin{center}
  \epsfxsize=15.0cm
 \epsfbox[10 10 560 290]{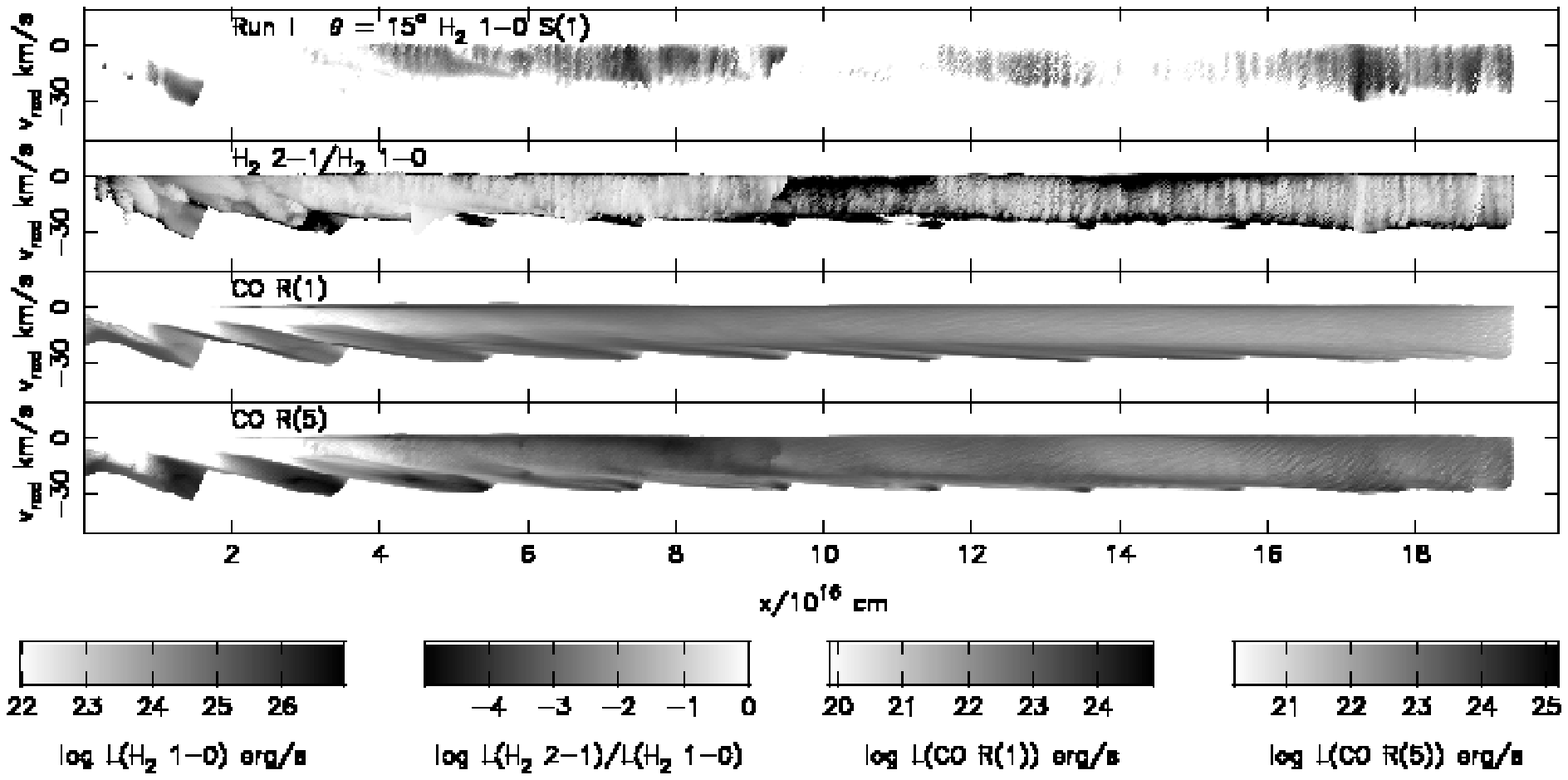}
\caption[]
{ Position velocity map for Run I in four molecular emission lines.  The jet axis is 15$^{\circ}$
out of plane of the sky toward the observer.  Each bin spans 2 $\times$ 10$^{14}$ cm in $x'$, the jet
axis projected on the plane of the sky, and 1 km~s$^{-1}$ in radial velocity. }
\label{posvel2}
  \end{center}
\end{figure*}

In the previous analysis for non-evolving molecular jets, the $\gamma$ determined for
CO was nearly uniformly larger than that for mass.  For the evolving jet simulations 
here, the dependence of CO luminosity on velocity is also usually steeper than the one 
for mass, but in some cases it is only equal (e.g., Run D at a viewing angle of 30$^{\circ}$) 
or, for one aborted jet simulation, was lower.  This occurred in a simulation of
an aborted jet that propagated into an equal pressure ambient medium.
The CO-derived $\gamma$ is lower than the mass-derived one at all viewing angles, and by as
much as 0.6.  However, in this simulation, the CO distribution can be better fitted with two power
laws, with the higher velocity one much higher than the single linear fit to the mass distribution.

Very little dependence of $\gamma$ on viewing angle is found for both of these simulation.
This is contrary to previous jet simulations (\citealt{1997A&A...323..223S},
\citealt{2001ApJ...557..429L}, \citealt{dc03}, and \citealt{rs03}),
which have shown an inverse relationship between viewing angle and
$\gamma$.  However, the CO-derived slopes in the high velocity range
are again an exception.  This group disagrees more strongly with the previous
results and, for small viewing angles, $\gamma$ varies in direct proportion to  the viewing angle.

An excess at the highest velocities is found  in Simulation I, as also
found in  previous studies of non-evolving molecular jets.
This bump  peaks at log~$v$/km~s$^{-1}$ = 1.4 (for
a viewing angle of 15$^{\circ}$) and is associated with the internal shocks from the pulsed jet.
In the aborted jet simulations, both the internal shocks and the high velocity bump are absent.

Recent observations of H$_2$ emission from protostellar outflows have been
similarly analysed \citep{2002ApJ...572..227S}.   As with the CO data, the H$_2$ 
flux distributions show a break in behaviour, with the break
between 2--17 km s$^{-1}$ for the selection of sources.
A flat or slightly rising flux is found at low velocities and a fall-off in 
flux with $\gamma$ in the range between 1.8 and 2.6 at high velocities.

\begin{figure}
  \begin{center}
  \epsfxsize=8.0cm
    \epsfbox[10 10 180 170]{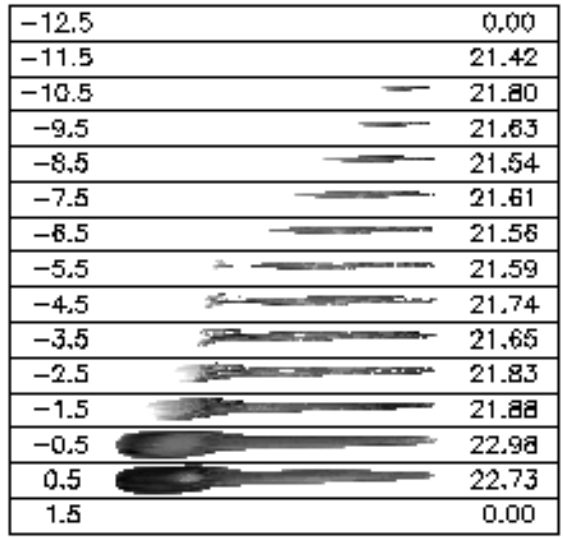}
\caption[]
{Velocity channel map for Simulation D in CO R(1) emission.   The jet is viewed 15$^{\circ}$ out
of the plane of the sky toward the observer. The central radial velocity in km~s$^{-1}$ is indicated
on the left side of each panel.  Four orders of magnitude in luminosity are displayed in each panel.
The log of the maximum luminosity in ergs~s$^{-1}$ is shown on the right side of each panel.
Each bin spans 2 $\times$ 10$^{14}$ cm in $x'$, the jet axis projected onto the
plane of the sky, and in $y$. }
\label{chan1}
  \end{center}
\end{figure}

From Fig.~\ref{gamma}, we see that Run I may best match these observational constraints. 
However, the H$_2$ flux distribution in Run I rises one
order of magnitude in brightness from the first H$_2$ datum in the figure to where 
it turns over near $v$ = 10 km~s$^{-1}$.
This is much larger than that found in the observations,
although it is a smaller rise than found in the non-evolving molecular jets simulations.
Additionally, the best fit line between $\log v$ = 1.0 and 1.3 has $\gamma$ =  -1.6, which
is in line with the observations.  This could be an adequate explanation for
sources viewed with their jet axis near the plane of the sky, such as
HH\,212.  However, at larger viewing angles (Fig.~\ref{gamma}
shows a viewing angle of 15$^{\circ}$ only), the low-velocity rise steepens.
In addition, at intermediate velocities the comparison becomes less good as the
viewing angle is increased. C-type shock physics may solve these
problems since ambipolar diffusion would generate considerably more emission near 
zero radial velocity in the absence of discontinuous acceleration.
The effect this has on the global properties of an outflow remain to be calculated.

\subsection{Position-Velocity Maps} 

We display position-velocity maps in H$_2$ and CO emission lines for Simulation D 
(Fig.~\ref{posvel1}) and Simulation I (Fig.~\ref{posvel2}).  The H$_2$ 1--0
position-velocity map for Simulation D shows emission only at the leading edge of the 
jet.  This emission is spread over 10 km~s$^{-1}$ and 1 $\times$ 10$^{16}$ cm along the projected
jet axis and is characterised by two peaks.  Note that steady state bow shock models
(see Fig.~14b of \citealt{skd03}) predict two peaks, which correspond to (1) the projected edge-on
leading edge near zero radial velocity and (2) the blue-shifted emission from the warm material projected
well behind the bow apex.

The pulsed jet in Simulation I shows intermittent H$_2$ 1--0 and 2--1 emission along
its length, with many emission sites spreading out over a large range of velocities.   
The H$_2$ 2--1 map shows that the sites of bright H$_2$ 1--0 emission
are also bright in the more highly excited line.
 
The emission in the CO lines shows a pair of ``Hubble-law" like dependences for 
Simulation D, but disconnected from the nozzle. The outer envelope of the first increases 
relatively quickly with position, and is followed by a shallower increase for much 
of the remainder  of the jet. While the emission in the CO R(1) line is nearly uniform
for non-zero velocities, there is some structure in the CO R(5) position-velocity
diagram. The bright spot near $x'$ = 7 $\times$ 10$^{16}$ cm is from the advancing edge 
of the shoulder region seen in Fig.~\ref{int1}, and the peak at $x'$ = 5 $\times$ 
10$^{16}$~cm is from the trailing filamentary structure.  Both of these bright features 
have radial speeds near zero.

For Simulation I, the CO R(5) diagram shows more structure than the diagram for CO R(1). 
Both show a prominent series of triangular shapes that distinguish the Hubble Law 
behaviour of a pulsed jet. There is evidence for a connective loop of emission between 
successive Hubble Law regions, until the triangles merge downstream, filling in all 
velocities, and the Hubble Law is then hidden.
The maximum luminosity in the CO R(5) map falls in between those of the H$_2$ 1--0
and the CO R(1) maps. Naturally, all of the maxima are higher than the corresponding
values in the aborted jet simulation.

\begin{figure}
  \begin{center}
  \epsfxsize=8.0cm
    \epsfbox[10 10 180 400]{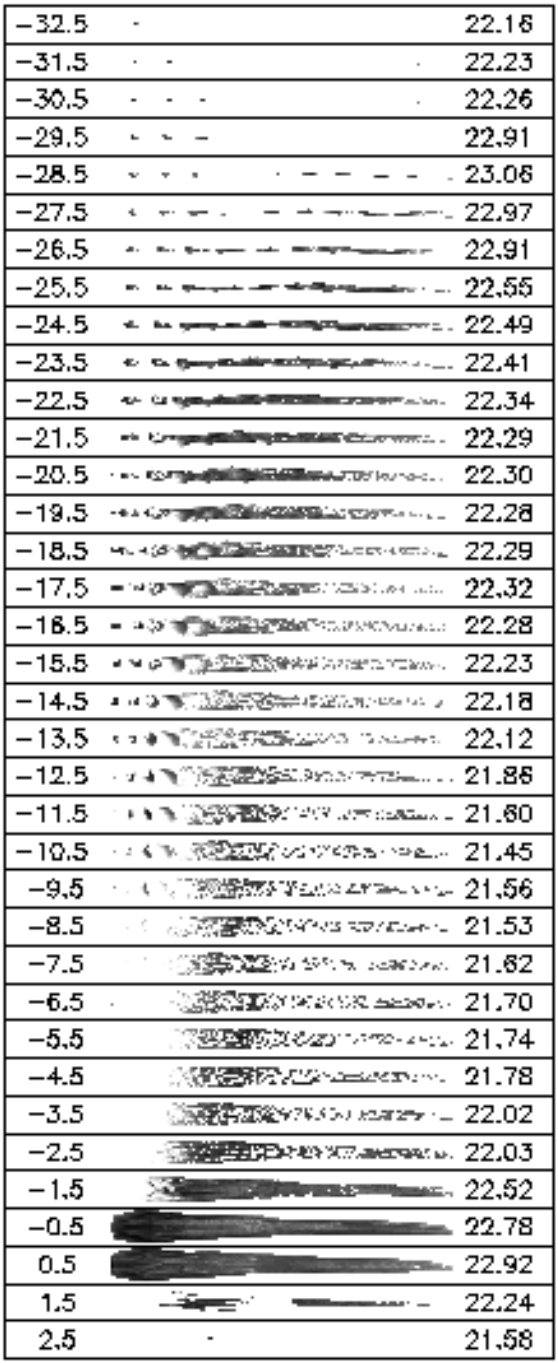}
\caption[]
{Velocity channel map for Simulation I in CO R(1) emission.  The jet is viewed 15$^{\circ}$ 
out of the plane of the sky toward the observer.  The central radial velocity in km~s$^{-1}$ is indicated
on the left side of each panel.  Four orders of magnitude in luminosity are displayed in each panel.
The log of the maximum luminosity in ergs~s$^{-1}$ is shown on the right side of each panel. Each bin
spans 2 $\times$ 10$^{14}$ cm in $x'$, the jet axis projected onto the plane of the sky, and in $y$. } \label{chan2}
  \end{center}
\end{figure}

\subsection{Velocity Channel Maps} 

We display CO R(1) velocity channel maps for Simulation D in Fig.~\ref{chan1}
and Simulation I in Fig.~\ref{chan2} with a resolution of 1 km~s$^{-1}$.
While the CO emission from Simulation D does not span a large
velocity range, the morphology does change as one moves from relatively
high radial velocity (10 km~s$^{-1}$) to low radial velocity.  At high radial velocities, only
the leading edge of material is visible, and as the velocity is decreased more emission appears closer
to the inlet. This is another aspect of the Hubble Law profile seen in the position-velocity 
map. Near zero velocity the entire envelope of the jet and cocoon of shock ambient material
is represented, along with the shoulder region seen in the density cross sections and
integrated CO maps.  This defines a T-type morphology. Even the filamentary structure 
seen in the integrated emission map for CO R(5) is seen to some extent in these
CO R(1) velocity channel maps.

The CO R(1) velocity channel maps for Simulation I reveal morphologies in the different 
velocity ranges more typical of our previous non-evolving jets \citep{rs03}. At high radial velocities,
only the internal shocks from the pulsed jet are seen and at low velocities the entire 
envelope of jet plus cocoon material is present.  At low velocities, the gap between the 
nozzle and the onset of emission decreases and a V-type global morphology is
identifiable. At intermediate velocities, a spinal structure is present.

\begin{figure}
  \begin{center}
  \epsfxsize=4.0cm
\epsfbox[10 10 160 60]{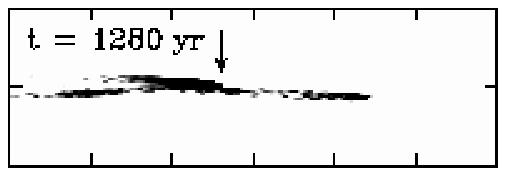}
  \epsfxsize=4.0cm
\epsfbox[10 10 160 60]{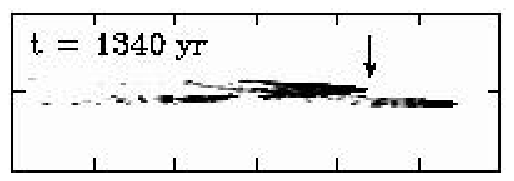}
  \epsfxsize=4.0cm
\epsfbox[10 10 160 60]{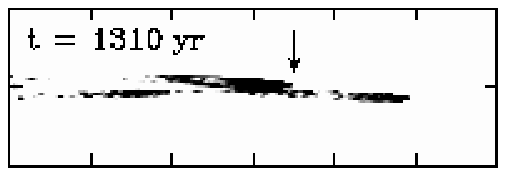}
  \epsfxsize=4.0cm
\epsfbox[10 10 160 60]{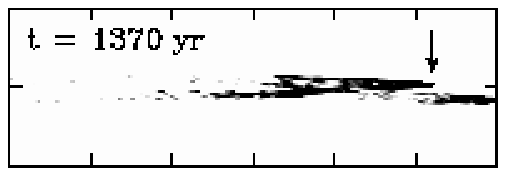}
\caption[]
{Birfucating bow shocks shown in the H$_2$ 1--0 line for Simulation I.  We display the portion of the
computational grid with $x$ = 1.4--2.0 $\times$ 10$^{17}$ cm. Adjacent tick marks are 1 $\times$ 10$^{16}$ cm apart.
An arrow points out the progress of the bow shock associated with the denser flow. }
\label{bifur}
  \end{center}
\end{figure}

The twin protuberances at the front of the jet are from the differing advance speeds
associated with the extrema of the evolving jet density in Simulation I.  As confirmed 
from a sequence of integrated H$_2$ 1--0 emission line maps (Fig.~\ref{bifur}),
a faster-moving bow shock leaves the path of the earlier slower-moving initial bow shock.  
This departure occurs at a fairly late time in the evolution of the jet ($t$ = 1300 yr, or
550 yr after the onset of the increase in jet density at the inlet) and well into the 
refocused leading part of the bow ($x$ = 1.6 $\times$ 10$^{17}$ cm).  The average speed
of the new bow is therefore close to 100 km~s$^{-1}$. An average speed for the propagation 
of the new bow shock equal to that of the (time-averaged) jet speed is not unreasonable 
in this simulation with the very light initial jet initially evacuating the path and with 
an acceleration from the refocused region.

\section{Application: an evolutionary diagnostic?} 
\label{application}

The structures and features detailed above can provide clues to
distinguish the evolutionary phase of a specific outflow. Statistically,
however, the integrated properties can provide the most employable
measures. In particular, the ratio L(H$_2$~1--0~S(1))/L(CO~R(1))
can be determined from simulations and compared to the observationally
derived L(H$_2$)/L$_{mech}$. The total molecular hydrogen emission
L(H$_2$) can be estimated from  L(H$_2$~1--0~S(1))
\citep{1995A&A...296..789S}, provided the near-infrared extinction can be
determined. The mechanical luminosity for both lobes in a bipolar outflow is derived from formulae of the
form L$_{mech}$ = M$_{co}$ $v_{co}^3$/(2\,$l$) where M$_{co}$ is
derived from L(CO~R(1)), assuming a CO abundance and temperature,
$v_{co}$ represents the appropriately averaged  speed of the mass that
dominates the outflow energy (and must account for the source orientation
when derived) and $l$ is the outflow size.

From the ratio of mass to L$_{co}$ in the log $v$ bins of Figure \ref{gamma}, we estimate from the simulations
that M$_{CO}$ for one side of a bipolar outflow $\sim$ 300~L(CO~R(1)) in solar units.
Furthermore, given the consistently low values of $\gamma$, we take  $v_{co}$ = 20~km~s$^{-1}$.
Finally, we fix L(H$_2$~1--0~S(1)/L(H$_2$) = 0.05. It must be remarked that a rather
high number of assumptions precede a comparison of
simulation and observation. Nevertheless, it is clear that
the ratio L(H$_2$~1--0~S(1))/L(CO~R(1)) $\sim$ 3--8 for Run D after
1000 yr, while  L(H$_2$~1--0~S(1))/L(CO~R(1)) $\sim$ 30--40 for Run I after
1000 yr. Given the above factors these convert to:
 L(H$_2$)/L$_{mech}$ $\sim$  0.02--0.05 for Run D
and  $\sim$  0.2--0.3 for Run I.

We confirm that strong H$_2$ flows are thus associated with Run I. For example,
we find L(H$_2$)/L$_{mech}$ $\sim$ 0.3 for Cepheus E, one of the dynamically 
youngest outflows \citep{sfe03}. 
Moreover, the H$_2$ emission is widespread and the CO position-velocity
diagram  displays structure cutting through all radial velocities at
all positions, rather than the triangular Hubble-law features
\citep{1997ApJ...474..749L}, consistent with the Run I predictions. 
Finally, we note that the CO $\gamma$ values for Cepheus E are 0.99 and 1.8,
for the two lobes, although these values are quite sensitive to the 
chosen local standard of rest \citep{1997A&A...323..223S}. Hence, observations of Cepheus E
are consistent with our results for a newly-forming outflow.

To date, CO emission has not been fully resolved in many outflows.
Velocity channel maps are  thus rare. A highly collimated CO outflow
emanates from IRAS~21391+5802, with a V-shaped morphology on velocity 
channel images \citep{2002ApJ...573..246B}. The V-shape suggests that 
it is also an immature outflow in which the
central source is increasing in power. We would thus expect to find a
strong H$_2$ outflow, as is, indeed, the case: the total shocked emission
is estimated to be 4\,L$_\odot$ whereas the mechanical power estimates lie
in the range 0.15--1.2\,L$_\odot$ \citep{2001A&A...376..553N}.

In contrast, there are many examples of outflows with very weak total
H$_2$ emission. These include the HH\,366/B5\,IRS1 outflow in the Perseus
cloud complex, which possesses high collimation. Hubble-law behaviour,
consistent with a single outburst
in the past is also present \citep{1999AJ....118.2940Y}. The 
H outflow in OMC-2 is another example, where the presently-observed H$_2$ jet does
not appear to be the driver for the CO outflow  \citep{2000AJ....120.1974Y}.

IRAS~18148-0440 in the L483 core is thought to be a protostar undergoing the 
transition from Class 0 to Class 1. Its CO outflow may possess a T-shaped morphology,
although background cloud subtraction makes this uncertain \citep{2000A&A...359..967T}. 
It belongs to a class of objects characterised by low velocity CO and no CO bullets, as
also predicted here for a transitional outflow \citep{2000A&A...359..967T}.

\section{Summary} 

We have reported on two jet simulations that model drastic jet density 
evolution over tens or hundreds of years.  The basic physical structure is 
similar to that previously found for non-evolving sheared molecular jets 
\citep{rs03}, with an accelerated leading bow shock followed by a wider,
slower shoulder-like region.   More specifically, both of the simulations 
on display here resemble a light molecular jet simulation.  The primary 
reason for this is that the shoulder region makes only slow
progress relative to the leading edge of the bow shock.

We have explored predicted features relevant to high resolution
near-infrared (H$_2$ rovibrational at 2.12$\mu$m emission), far-infrared
(CO R(5) rotational emission at 435$\mu$m) and millimetre
(CO R(1) rotational emission at 1.3mm) wavelength observations.
As expected, the H$_2$ lines reveal patchy details of the structure and 
the CO lines show the general outline of material encompassed by the bow.  
CO line emission from mid-level rotational states show both the general
outline and the internal shock structure.  Owing to the small amount of
higher temperature gas in the aborted jet simulation, the H$_2$ emission 
is limited to the very tip of the accelerated bow shock. The predicted 
values of the 1--0~S(1) 2.12$\mu$m luminosity are comparable to those 
derived for the majority of outflows in the unbiased statistical study 
of \citet{2002A&A...392..239S}.

The evolution in the integrated luminosity in these molecular lines is 
evident in both simulations.  The changes in the integrated values of 
H$_2$ occur at program times that are within 100 yr after the 
modification of the inflow jet density. Of particular note is the ratio 
of the integrated H$_2$ to the integrated CO R(1) line in the aborted 
jet case, which decreases dramatically with time.  Thus, in principle 
this ratio can be used to estimate the age of an outburst or aborted 
molecular flow from a protostellar core.

We have also computed velocity distributions in mass, CO luminosity, 
and H$_2$ luminosity.  Here the main results are:

1.  The slopes of these distributions are small, closer to one than two 
in most cases. Steeper slopes ($\gamma$ = 2--3) are seen at high 
velocities in CO in the aborted jet simulation.

2. The slopes for the aborted jet case are larger than those for the 
increasing jet density case, although not by a large amount for the 
simulations listed in Table \ref{gammas}.

3. Similar to non-evolving jets \citep{rs03}, $\gamma_{\rm CO}$ is
usually greater than $\gamma_{\rm mass}$, but there are significant 
exceptions.

4. In the aborted jet simulation, there is no excess in the amount of mass
or CO at the highest velocities; we have seen this excess that is associated 
with the internal shocks of a pulsed jet.

The position-velocity maps for the aborted jet case are relatively
featureless.   In CO emission, a triangular Hubble-Law outline is
filled in by the emission but does not include the low velocity emission
near the jet inflow boundary.  The CO R(5) position-velocity map
shows some fine structure, with brighter regions associated with
the warmer parts of the flow: the leading part of the bow shock and
some internal structure near the front of the shoulder region.

In the increasing jet density simulation, the position-velocity CO maps
show the series of Hubble Law regions associated with successive
pulses in the jet.  Additionally, the CO R(5) map has more structure
than the CO R(1) map.  In the H$_2$ position-velocity map, there are numerous
features that are narrow along the jet axis but cover a large range
of velocities (the vertical lines in Fig.~\ref{posvel2}).

The CO R(1) velocity channel maps for the aborted jet simulation have
emission confined to a small range of velocities.  The sequence of
channel images shows a T-morphology at low velocities and the
characteristics typical of a Hubble Law at high velocities.  The channel
maps for the increasing jet density simulation possess a
V-morphology, with the inclusion of an additional protuberance
from the new denser flow that carves a fresh path near the previous flow.

These features, combined with the levels of integrated emissions,
suggest that jets in certain outflows (e.g. IRAS~21391+5802
and Cepheus E, see Sect.~\ref{application}) are increasing
in power, while other jets (e.g. that from IRAS~18148-0440 in L483)
are on the decline.

Not many detailed velocity channel maps have as yet reached the literature.
This will change shortly, and with the advent of new telescopes such as
SIRTF, FIRST and ALMA, imaging and spectroscopy will reveal the properties
on fine scales in many molecular lines of intermediate and low
excitation, such as in the  CO R(5) and R(1) lines. Here we have
determined how these properties can help determine the jet evolutionary stage.

\section*{Acknowledgments}

The numerical calculations were run on the local SGI Origin 2000 computer (FORGE),
acquired through the PPARC JREI initiative with SGI participation.
AR is funded by PPARC.

\bibliography{jetbib}

\label{lastpage}

\end{document}